\documentclass[]{article}

\usepackage{amsmath}
\usepackage{amssymb}
\usepackage{authblk}
\usepackage{bm}
\usepackage{graphicx}
\usepackage{hyperref}
\usepackage[square,numbers]{natbib}
\usepackage{times}
\usepackage{url}

\DeclareMathOperator\erf{erf}

\begin{document}

\title{Continuous Flow Model of a Historical Battle:\\ A Fresh Look at Pickett’s Charge}

\author[1]{Jonathan Poggie}
\author[2]{Sorin A. Matei}
\author[3]{Robert Kirchubel}

\affil[1]{School of Aeronautics and Astronautics, Purdue University, jpoggie@purdue.edu}
\affil[2]{FORCES Initiative, College of Liberal Arts, Purdue University, smatei@purdue.edu}
\affil[3]{FORCES Initiative, College of Liberal Arts, Purdue University, rkirchub@purdue.edu}

\renewcommand\Affilfont{\itshape\small}

\maketitle

\begin{abstract}
  A continuous flow model of infantry behavior, based on conservation
  of individuals and tracking of subunit identity, has been developed
  in sufficient detail that it can now be applied to a realistic
  simulation of a historical battle. Pickett’s charge during the 1863
  Battle of Gettysburg, Pennsylvania in the U.S. Civil War was chosen
  as an initial application of the model.  This scenario is a good
  test of the current mathematical model because many modern military
  tactics were employed, in a context where the action took place on
  foot or horseback, and the historical map and troop numbers are
  available.  Compared to a discrete agent model, the flow model was
  found to better capture the interaction of the forces with the
  terrain and each other.  A brigade-level simulation, faithful to the
  details of the historical events, was performed.  The main source of
  asymmetry in the numbers of casualties was found to be the inability
  of the Confederate forces to use effective ranged fire while they
  were moving. Comparison of simulations with and without terrain
  effects showed that they slow the pace of battle and favor the
  defenders, exposing the attackers to heavy ranged fire for an
  extended period. A statistical analysis of possible outcomes for an
  ensemble of 1000 randomized perturbations of the baseline
  brigade-level scenario was carried out. Consistent with historical
  events, it was found that only 6\% of the scenarios resulted in an
  outcome that could be considered a Confederate victory.
\end{abstract}

\section{Introduction}

Conventional military simulations (wargames) and related game-like
simulations typically use discrete agents, essentially virtual game
pieces, to model groups of individuals and machinery. This aggregation
of discrete entities into groups can be very arbitrary, and it may
discard much complex and important individual behavior.

Continuous agents (probability densities of individuals) could
potentially overcome these difficulties. A continuous flow model could
more realistically portray group motion, provide decision makers with
a more intuitive understanding of the momentum of a conflict, and
become a useful tool for training and decision making.

These ideas have been explored to a limited extent in previous
literature. An early effort to model discrete entities as a flow was
made by \citet{lighthill55}, who developed a fluid dynamic model of
traffic flow. \citet{protopopescu89} and \citet{fields93} extended
these ideas to combat modeling. \citet{hughes02} adapted the
\citet{lighthill55} model to human crowds, and later applied the
approach to modeling the 1415 Battle of Agincourt \citep{clements04}.
\citet{spradlin07} examined different kinds of \citet{lanchester14}
fire models in the context of continuous units. Adapting swarming and
schooling models from mathematical biology, \citet{keane11}
implemented non-local effects through convolution integrals in order
to enforce unit formations. \citet{gonzalez11} created an advanced
version of a reaction-diffusion model \citep{fields93}, in which
motion responsive to enemy force concentration was implemented.

In the context of infantry combat, the models proposed in these
previous studies over-emphasize the fluid aspects of unit motion, and
they lack an efficient means to represent tight infantry
formations. In \citet{poggie20} we introduced a new continuous flow
model of infantry behavior in a military conflict. The model
incorporated conservation of individuals, effects of crowd density and
surface inclination on walking speed, and \citet{lanchester14} combat
models for both close-in and ranged fire. A new approach to tracking
subunit identity was introduced to implement unit movement and
orientation to better represent infantry formations.

In the present work, we have brought in refinements to the model,
including discrete artillery units, an overlay layer to represent
different kinds of terrain on the map (vegetation, fences, buildings,
etc.), improvements to movement and ranged fire, a list of orders for
each unit, unit morale, and checks for retreat and pressing an attack.

With this new approach, a battle flow model can be used for efficient
and realistic simulations. We illustrate this point by applying the
method to a historical battle.

\subsection{Application to Pickett's Charge}

As an initial historical application of the model, we chose the Battle
of Gettysburg, Pennsylvania in the U.S. Civil War, focusing on the
famous Pickett’s charge of July~3, 1863. This application is a good
test of our current mathematical model because many modern military
tactics were employed in a context where the action took place on foot
or horseback.

Furthermore, the details of the historical battle are relatively well
known.  Maps and numbers for soldiers and equipment are available from
historical records.  Maps in \citet{gottfried07} and \citet{laino14}
provide a reconstruction of the timeline and unit motion through the
battle.  \citet{busey05} have published a comprehensive tabulation of
the forces present.

The American Civil War (1861--1865) pitted the agricultural South
(Confederacy, Red) against the industrial North (Union, Blue).  In a
dispute that arose over whether newly acquired western territories
would allow slavery, Southern states seceded from the United States
starting in December 1860 \citep{weigley00}. Fighting began at Fort
Sumter, South Carolina in April 1861. Union successes in 1864
eventually led to Confederate surrender in 1865, with a cost of at
least million combined casualties (about 3\% of the
population). Nonetheless, at the time of the Battle of Gettysburg
(July~1--3, 1863), the final outcome of the war was very much in
doubt.

In spring 1863, the Confederate forces initiated an invasion of
southern Pennsylvania.  The Union army moved to position itself
between the invaders and the threatened Washington, DC.  The two sides
met at the crossroads of Gettysburg, PA.

The Battle of Gettysburg was the largest battle ever fought in North
America.  On the first two days of fighting (July~1--2), each army
suffered tremendous losses with inconclusive results. On the third day
(July~3), the armies were positioned on opposite ridgelines known as
Seminary Ridge and Cemetery Ridge. The ridges ran approximately
north-south, and were separated by a flat valley about 1~km wide.  The
valley was occupied by several farms, with fenced fields, drainage
ditches, and a road running north-south.

At 13:00 on July~3, the Confederate artillery opened fire, and the
Union guns replied. The Union artillery commander soon silenced his
guns to conserve ammunition. The Confederate artillery continued to
fire, using up much of its ammunition \citep{gallagher94,reardon16}.

Sometime after 14:00, Confederate infantry left the shelter of
Seminary Ridge and began a frontal assault. (A simplified map of the
situation is given in Figure~\ref{fig:map}.)  Union guns opened fire
across the valley. After marching a kilometer under large caliber
shelling, the advancing Confederate infantry reached the fence-lined
Emmitsburg Road. They had to stop to tear down the fences, under
bombardment, and when they got moving again, they did so in a
staggered fashion without their earlier cohesion. (We replicate this
behavior in the model, as described later.) Additional Union guns to
the north and south turned on the attacking mass, enfilading it.

With the enemy approximately 300~m away, and now properly charging
(captured in the model as faster forward motion), the sheltered Union
infantry opened fire. The Union command continued to send
reinforcements, and even had units to the left and right advance
slightly, adding more enfilading fire. These effects combined to
funnel the Confederate attack into a narrower, denser mass. Since
infantry of that era could not simultaneously fire and move,
Confederate troops did not fire until within about 100~m of the
defenders.

Small groups of Southern soldiers reached Union lines, notably near
the now-famous copse near the center of the Union lines. Just to the
north, at The Angle, a few broke through the Northern line. Each time,
Union troops rallied and threw the attackers back. By 15:30, the
assault petered out, and disorganized attackers withdrew.

\subsection{Previous Simulations}

A few previous wargame-type studies have addressed the Battle of
Gettysburg. \citet{tillman96} examined Ewell’s decision not to attack
Culp’s Hill on July~1, 1863, using the Janus combat simulation
software. That software provides stochastic simulation of combat in
three dimensions and time. It is a discrete agent model, and each
entity in this specific simulation represented either 80 infantrymen
or 2 artillery pieces. Comparing four alternative scenarios, they
concluded that Ewell’s decision was correct: he did not have the
forces to take the hill, and even with plausible additional forces he
could not have held the hill for long.

\citet{armstrong15} modeled Pickett’s Charge using the Lanchester
aimed fire equations (square law), testing sensitivity of the outcome
to changes in troop strength within the historical uncertainty and
changes in Confederate tactics. There is no map or terrain in their
zero dimensional model, only a phased interaction (artillery barrage,
Confederate advance, skirmish) of aggregated units. They concluded
that the Confederates could plausibly have taken the Union position
under a different tactical approach, but that the Confederates lacked
sufficient forces to exploit such a victory. They write: “Thus, it
seems that while Lee could have captured The Angle via a better
barrage and a larger charge, adding more men to the initial assault
would have added significant risk to his ability to exploit a
breakthrough and achieve the larger victory he was seeking.”

With the use of a model based on continuous densities of individuals,
our approach differs from these previous discrete agent
simulations. Our results agree in some ways with \citet{armstrong15},
but in including the effects of terrain and unit flow, and
particularly the interaction with fences, we capture new behavior in
the model and come to a slightly different conclusion.

\section{Mathematical Model}
\label{sec:math_model}

The mathematical model used in the present work is an extension of the
continuous flow model introduced in \citet{poggie20} The basis of the
model lies in conservation of individuals and tracking of subunit
identity. The pattern of flow of a unit across the map is a result of
the interaction of impediments of the environment with the unit's
efforts to maintain formation in following its ordered path.

\subsection{Unit Motion}

A fundamental assumption of the model is that individuals can be
tracked; they do not spontaneously appear or disappear. Individuals
are aggregated into units (here brigades or armies), which will be
enumerated with the subscript $i = 1, 2, \dots, N$. The local density
of members of each unit is governed by a conservation equation
\citep{hughes02} of the form:
\begin{equation}
  \label{eq:continuity}
  \frac{\partial \rho_i}{\partial t}
  + \nabla \! \cdot \! \left( \rho_i \boldsymbol{u}_i \right)
  = - \omega_i
\end{equation}
where t is time, $\nabla$ is the spatial ($\boldsymbol{x}$) gradient
operator, the density of individuals per area for Unit-$i$ is $\rho_i$,
the walking velocity is $\boldsymbol{u}_i$, and the rate of casualties
per area is $\omega_i$.

The velocity consists of a combination of directed motion and
diffusion:
\begin{equation}
  \label{eq:velocity}
  \boldsymbol{u}_i = \boldsymbol{V}_i - \frac{D_i}{\rho_i} \nabla \rho_i
\end{equation}
For the present work, diffusion is retained as an option in the
computer code, but the diffusion coefficient has been set to $D_i=0$
because it did not seem to contribute to the realism of the
simulation. (The military tactics of the Civil War era involved tight
formations; diffusion tends to make a formation gradually spread out.)
The directed motion is affected by local crowd density, ground slope,
proximity to goal location, terrain type, and unit orders, summarized
as:
\begin{equation}
  \label{eq:velocity_directed}
  \boldsymbol{V}_i = V_m \, f(\rho) \, g(s) \, h(r) \, T(\boldsymbol{x} ) \, \boldsymbol{d}_i
\end{equation}
Here $V_m$ is the maximum walking speed, $\rho = \sum_i \rho_i$ is the
local total density, $\boldsymbol{d}_i$ is the walking direction,
$s = \boldsymbol{d}_i \! \cdot \! \nabla H$ is the ground slope in the
direction of walking (directional derivative), and r is the distance
from the destination. The variable $H(\boldsymbol{x})$ is the local
elevation, obtained from a topographical map of the area of interest.
The variable $T(\boldsymbol{x})$ represents the effect of local
terrain type.

Speed is assumed to slow with increasing crowd density according to
the formula:
\begin{equation}
  \label{eq:velocity_density_effect}
  f(\rho)
  = -6 \left( \frac{\rho}{\rho_m} \right)^5
  + 15 \left( \frac{\rho}{\rho_m} \right)^4
  - 10 \left( \frac{\rho}{\rho_m} \right)^3
  +1
\end{equation}
The values for the parameters in
Eqs.~\eqref{eq:velocity_directed}--\eqref{eq:velocity_density_effect},
$V_m=1.4$~m/s and $\rho_m = 5.6$~m$^{-2}$, are based on data on
pedestrian walking speed under crowded conditions
\citep{greenshields35}.  Equation~\eqref{eq:velocity_density_effect}
is chosen to obtain a smooth transition from walking at full speed,
$f ( \rho ) = 1$ and $V=V_m$, under very low density of individuals,
to no motion, $f ( \rho ) = 0$ and $V=0$, at a maximal crowd density
of $\rho_m$. A discussion of some of the possible alternative options
for this function is given in \citet{poggie20}.

Given the relatively gentle slope of the Gettysburg terrain, walking
speed is assumed to be affected by terrain according to the following
simple formula:
\begin{equation}
  \label{eq:velocity_slope}
  g(s) =
  \begin{cases}
    \frac{1}{1 + 3.5 \, s}&  \text{if $s > 0$ (uphill)},\\
    1&  \text{if $s \le 0$ (downhill)}.
  \end{cases}
\end{equation}
This model assumes that walking downhill does not significantly affect
speed. Additional discussion of this function is given in
\citet{poggie20}.

Fences, roads, buildings, and local vegetation can affect the speed of
infantry motion. This effect is accounted for by a map overlay layer,
denoted by the function $T(\boldsymbol{x} )$. The overlay function is
intended not only to represent physical impediments to motion, but
also to account for psychological effects of an obstructed field of
view and small sub-grid scale motions that are not resolved in the
simulation. The elevation $H(\boldsymbol{x})$ and terrain
$T(\boldsymbol{x})$ fields for the Gettysburg battlefield will be
discussed later, in the context of Figure~\ref{fig:map}.

Unit formations are implemented through a subunit identity function
\citep{poggie20}. The identity of a subunit is tracked using its
position in the initial formation. The following equation tracks the
origin of a subunit $\boldsymbol{\xi}_i$:
\begin{equation}
  \label{eq:subunit_identity}
  \frac{D \boldsymbol{\xi}_i}{D t}
  = \frac{\partial \boldsymbol{\xi}_i}{\partial t}
  + \left( \boldsymbol{u}_i \! \cdot \! \nabla \right) \boldsymbol{\xi}_i
  = 0
\end{equation}
Making an analogy to fluid mechanics, this formulation is equivalent
to setting the material derivative to zero; identity does not change
following the flow.

Self-aware subunits can have individual goals and position relative to
the unit’s centroid. To define the goals, we use a goal potential
related to the current distance from a goal position:
\begin{equation}
  \label{eq:goal_potential}
  \phi_i ( \boldsymbol{x} , \boldsymbol{\xi}_i )
  = \left| \boldsymbol{x} - \boldsymbol{z}_i ( \boldsymbol{\xi}_i ) \right|^2
\end{equation}
Here $\boldsymbol{x}$ is the current position and $\boldsymbol{z}_i$
is the desired position for a particular subunit identified by
$\boldsymbol{\xi}_i$. The subunit walks in the direction that
minimizes the goal potential:
\begin{equation}
  \label{eq:walking_direction}
  \boldsymbol{d}_i = - \frac{\nabla \phi_i}{ | \nabla \phi_i | }
\end{equation}

If a subunit is near the target location (about 3 grid cells), the
speed is reduced as an error function of distance. This allows each
part of the unit to eventually come to rest at a stable fixed
position. If a subunit is very far from the goal (about 40 cells), the
units march double time to catch up. This behavior is captured by the
function:
\begin{equation}
  \label{eq:goal_proximity}
  h ( r )
  = \erf \left( \frac{r}{1.5 \, \Delta s} \right)
  + \frac{1}{2} \left[ 1 + \erf \left( \frac{r - 40 \, \Delta s}{10 \, \Delta s} \right) \right]
\end{equation}
where $r = \left| \boldsymbol{x} - \boldsymbol{z}_i \right| $ and
$\Delta s$ is the grid (map) resolution.

We set the goal through translation and rotation from the initial
position:
\begin{gather}
  \label{eq:goal_position}
  \boldsymbol{z}_i = \boldsymbol{A} \left( \boldsymbol{\xi}_i - \boldsymbol{y}_{1,i} \right) + \boldsymbol{y}_{2,i}\\
  \label{eq:goal_matrix}
  \boldsymbol{A} =
  \begin{bmatrix}
    a (t) \cos \theta (t) & -b (t) \sin \theta (t) \\
    a (t) \sin \theta (t) & b (t) \cos \theta (t)
  \end{bmatrix}
\end{gather}
Here $\boldsymbol{y}_{1,i}$ represents the initial center for Unit-$i$
and $\boldsymbol{y}_{2,i}$ is the current target location of the
center. By varying $\boldsymbol{y}_{2,i}$ with time, the center of the
unit can be made to follow an arbitrary, curved path. The variable
$\theta$ establishes rotation; it too can vary with time. The
parameters a and b control stretching along the coordinate axes; these
allow the shape of the formation to change. Although the
transformation specifies rigid translation and rotation, the resulting
motion of the unit is fluid, not rigid. Each subunit will do its best
to reach the goal position, encountering different conditions along
the way.

\subsection{Combat Model}

Combat is divided into two categories: close-in combat and ranged
combat. Close-in combat is intended to represent hand-to-hand combat
and fire at a sufficiently close range that the attacker can pick an
individual target. Ranged fire represents fire at a sufficient range
that the target is the general enemy unit and not an individual. The
notation $\omega_i=\omega_i^{\prime}+\omega_i^{\prime \prime}$ will be
used to designate these two parts of the losses.

Close-in combat is modeled according to the Lanchester area fire model
\citep{lanchester14,mackay09}. The rate of fire is proportional to the
local concentration of attackers $\rho_j$. The number of available
targets is proportional to the local density of defenders
$\rho_i$. Summing the result over all enemy units, the loss rate for
Unit-i due to close-in combat is:
\begin{equation}
  \label{eq:combat_closein}
  \omega_i^\prime =\rho_i \sum_j k_{ij} \rho_j
\end{equation}
where $k_{ij}$ is a proportionality coefficient for each pair of
combatants. The summation could potentially also include friendly fire
($k_{ij} \ne 0$ for $i=j$). In general, the proportionality
coefficient is not necessarily symmetric ($k_{ij} \ne k_{ji}$); the
effectiveness of the fire of one unit is not necessarily the same as
another. Nonetheless, in the absence of detailed information on the
subject, we assume a baseline value of
$k_{ij} = 5.0 \times 10^{-2}$~m$^2$/s for all units, without friendly
fire ($k_{ij} = 0$ for $i=j$). This figure is on the same order as
those employed in previous studies with Lanchester models
\citep{bracken95}, and it has given plausible results in our tests. An
additional tapering of close-in combat for very small densities,
$\rho_i < 10^{-4} \rho_m$ was used to eliminate unrealistic combat in
regions where there were essentially no forces.

Several options could be considered for ranged fire, including a
highly detailed model in which each attacking subunit had a different
target subunit. Such a model is awkward and inefficient to implement
in a computer program employing domain decomposition to achieve
parallel speedup, primarily because the attacker and target may lie in
different domains, with their information held on separate computer
processes. The data would need to be exchanged via message
passing. For the present work, therefore, we chose to implement ranged
fire by treating all units as discrete, aggregated entities. Under
this model, only the data summarizing the status of the unit as a
whole need to be exchanged. An aggregated attack is applied to the
defender, and losses are spread across the defender’s constituents.

For ranged fire under this approach, each attacking unit must first
choose a target unit. Here we choose the nearest enemy centroid within
an angular window of $\pm 45^\circ$ from the current bearing of the
attacking unit.

The loss rate for Unit-$i$ due to ranged fire is given by:
\begin{equation}
  \label{eq:combat_ranged}
  \omega_i^{\prime \prime} = \rho_i \sum_j k_{ij}^{\prime}  \, f_1 \, f_2 \, f_3 \, f_4 \, f_5  \, R_j
\end{equation}
where $R_j = \iint \rho_j \, dA$ is the total number of elements or
individuals in each attacking force. For the brigade-level simulations
(Section~\ref{subsec:brigades}), the ranged fire efficiency for
infantry was taken to be $k_{ij}^\prime=8.0$~s$^{-1}$ and that of
artillery was $k_{ij}^\prime = 16.0$~s$^{-1}$.

Several factors are considered that affect the efficiency of ranged
fire: target direction, target range, attacker and defender bearing,
elevation difference, and attacker motion. The range function is:
\begin{equation}
  \label{eq:f1}
  f_1 ( r_{ij} ) =
    \begin{cases}
    1&  \text{if $r_{ij} < R_0$},\\
    \left( R_0 / r_{ij}\right)^2&  \text{if $r_{ij} \ge R_0$}.
  \end{cases}
\end{equation}
where $r_{ij}$ is the distance between the attacker and the
defender. The parameter $R_o$ is a characteristic range, taken to be
100~m for infantry rifle fire and 1200~m for artillery fire.  For
infantry units, firing to the front should be more effective than
firing to the side. The historical infantry units were trained and
organized to primarily fire to the front; getting a clear shot to the
side might be difficult. Considering the angle $\alpha$ between the
target direction and the attacker bearing (both angles as seen from
the attacker’s centroid), this effect is implemented with:
\begin{equation}
  \label{eq:f2}
  f_2 ( \alpha ) = \exp \left( -2 \, \alpha^2 / \alpha_r^2 \right)
\end{equation}
With this function, shooting straight ahead ($\alpha = 0^\circ$) is
effective, whereas shooting at $\alpha = \alpha_r$ is highly
ineffective. Here the characteristic angle parameter is taken to be
$\alpha_r = 90^\circ$.

Similarly, relative orientation should have an effect. Flanking fire
should be more effective than head-on fire. If $\beta$ is the angular
difference between defender bearing and attacker bearing (again, as
seen from the unit centroid), the effect of relative orientation is
taken as:
\begin{equation}
  \label{eq:f3}
  f_3 ( \beta ) = ( 3 + \cos \beta ) / 2
\end{equation}
Specifically, opposing units facing toward each other
($\beta = 180^\circ$) gives the baseline of $f_3=1.0$, flanking fire
($\beta = 90^\circ$) gives an improved effectiveness of $f_3=1.5$, and
attacking an enemy’s rear ($\beta = 0^\circ$) doubles aimed fire
effectiveness, $f_3=2.0$.

Shooting up or down should also diminish accuracy. Taking $\theta$ to
be the angle of fire relative to horizontal, we implement the effect
of elevation difference as:
\begin{equation}
  \label{eq:f4}
  f_4 ( \theta ) = \exp \left( -2 \, \theta^2 / \theta_r^2 \right)
\end{equation}
For the present work, the elevation reference angle is taken to be
$\theta_r = 30^\circ$. In future implementations of the model, we hope
to improve the realism of this effect by including a line-of-sight
check. That is, an improved model would check for terrain in the way
of a rifle shot, but perhaps not for an artillery shot.

A moving attacker should be less effective in ranged fire. The effect
of attacker speed $V$ is:
\begin{equation}
  \label{eq:f5}
  f_5 ( V ) = \exp \left( - 50 \, V / V_r \right)
\end{equation}
where $V_r=V_m$ is a reference speed. This model provides a strong
penalty for shooting while moving. Stationary defenders have a strong
advantage. This choice seems reasonable for Civil War era infantry
combat.

\subsection{Artillery}

Since the artillery units considered for the present model of
Pickett’s Charge consisted of at most 15 guns, the continuous flow
model did not seem appropriate. Thus, a mixed model was employed, with
special discrete units selected to represent artillery. These are
traditional wargame units.

Although the framework exists in our computer code for moving discrete
units, the artillery units were specified to be immobile on the time
scale of the simulation. Further, we did not allow losses for
artillery units, nor did they participate in close-in combat. Again,
these simplifications could be relaxed in future versions of the
model.

For the purposes of computing infantry motion around the artillery
units, the latter were assigned an effective density distribution. The
peak value was taken to be $0.1 \rho_m$ with Gaussian decay away from
the center with a characteristic scale of 20~m.

\subsection{Unit Orders and Breakpoints}

The computer code includes a facility to provide each unit with a list
of orders. For the Confederate units, there were three actions: rotate
to face the direction of intended motion, translate to bring the unit
centroid to a specified location, and rotate to face the nearest
enemy. For the Union units, the list of orders was to wait until a
Confederate unit was within 500~m, rotate, translate, and face the
enemy. For two Union units (combined 8~Ohio and 126~New York
regiments, and Stannard’s brigade), the orders were modified to flank
the charging Confederates with a combined translation and rotation
motion. (See Figure~\ref{fig:map} and Table~\ref{tbl:brigades} for the
units involved in the simulations.)

The algorithm attempted to have the units carry out their orders in a
smooth manner. In other words, $\boldsymbol{y}_{2,i}$ in
Eq.~\eqref{eq:goal_position} was set to attempt a slow march of
0.6~m/s and $\theta ( t )$ was set to attempt to complete a rotation
of the formation at a comparable rate. A tapering function slowed down
the ordered translation and rotation as the desired position and
formation were approached. Characteristic values of 10~m and
$10^\circ$ defined the onset of the tapering.

These variables represent goals for the unit as a whole; the subunits
are not necessarily able to keep up, given the constraints of terrain,
crowding, and so on. As shown in the examples, the result is an
irregularity in distribution of forces. In this way, the continuous
flow model adds realism over models based on discrete unit motion.

A provision to modify the orders to either retreat or press an attack
is also included. At the beginning of each time step, the current
fractional losses $F_i$ of each unit were calculated. A morale
variable was then computed as follows:
\begin{equation}
  \label{eq:morale}
	M_i = M_i^0 - F_i / F_{i,m}
\end{equation}
where the initial morale is $M_i^0$ and the reference fractional
losses were set to $F_{i,m}=0.35$.  Then, for each unit, a series of
checks was performed. First, the nearest enemy for the given unit was
found in a sector of $\pm 45^\circ$ from the unit’s bearing. If that
closest visible enemy was in retreat, the selected unit was given a
morale increment of 0.5. If a visible Union unit was retreating,
Confederate units with positive morale were ordered to press the
attack, pursuing the enemy beyond their original orders. For negative
morale, units of both sides withdrew to their initial positions.

\subsection{Numerical Implementation}

Simple boundary conditions, consistent with the mathematics of the
model, were used at the edge of the map. For flow in, the density
$\rho_i$ was set to a very small value and the identity
$\boldsymbol{\xi}_i$ was set to the local coordinates. For flow out,
the appropriate boundary conditions would be extrapolation of the
values near the boundary. Nonetheless, to avoid having units, or
portions of units, leave the map, the directed velocity component
normal to the boundary $\boldsymbol{V} \! \cdot \! \boldsymbol{n}$ was
tapered to zero around the border of the map. The width of this
boundary region was 2\% of the map dimension.

The main computer program was implemented in modern, object-oriented
Fortran. Spatial derivatives of the convective terms in
Eq.~\eqref{eq:continuity} and Eq.~\eqref{eq:subunit_identity} were
computed using simple, second-order upwind differencing based on the
directed velocity for the density equation and the total velocity for
the subunit identity equation. Employing a harmonic
limiter \citep{gaitonde93} worked well to avoid numerical
oscillations. Other spatial derivative terms were calculated with
second-order central differences. Time integration was carried out
using second-order Runge-Kutta time stepping.

Higher-order and implicit time stepping schemes were tested, as well
as central spatial differences with explicit numerical dissipation,
but these approaches did not appear to offer any significant
advantages over the methods selected here.

The computational cost of the model is relatively high compared to
discrete agent approaches with an equal number of units; the model is
highly efficient, however, compared to a discrete model with enough
agents to capture the subunit motion. To achieve a reasonable run time
(in terms of wall clock), multi-level parallel calculation methods
were employed. At the large scale, domain decomposition was
employed. The map was split into overlapping tiles, with each tile
assigned to a computational process. Communication of data in the
overlapping boundary regions (fringe cells or ghost cells) was
implemented using the MPI library. Each tile corresponded to an MPI
rank, and messages were exchanged using persistent sends and receives.

At a finer level of parallelism, outer loops were multi-threaded using
the OpenMP library. Inner loops were automatically vectorized by the
compiler (tested with Intel ifort and GNU gfortran).

\section{Pickett's Charge Scenario}

\begin{figure}
  \centering
  \includegraphics[width=0.5\linewidth]{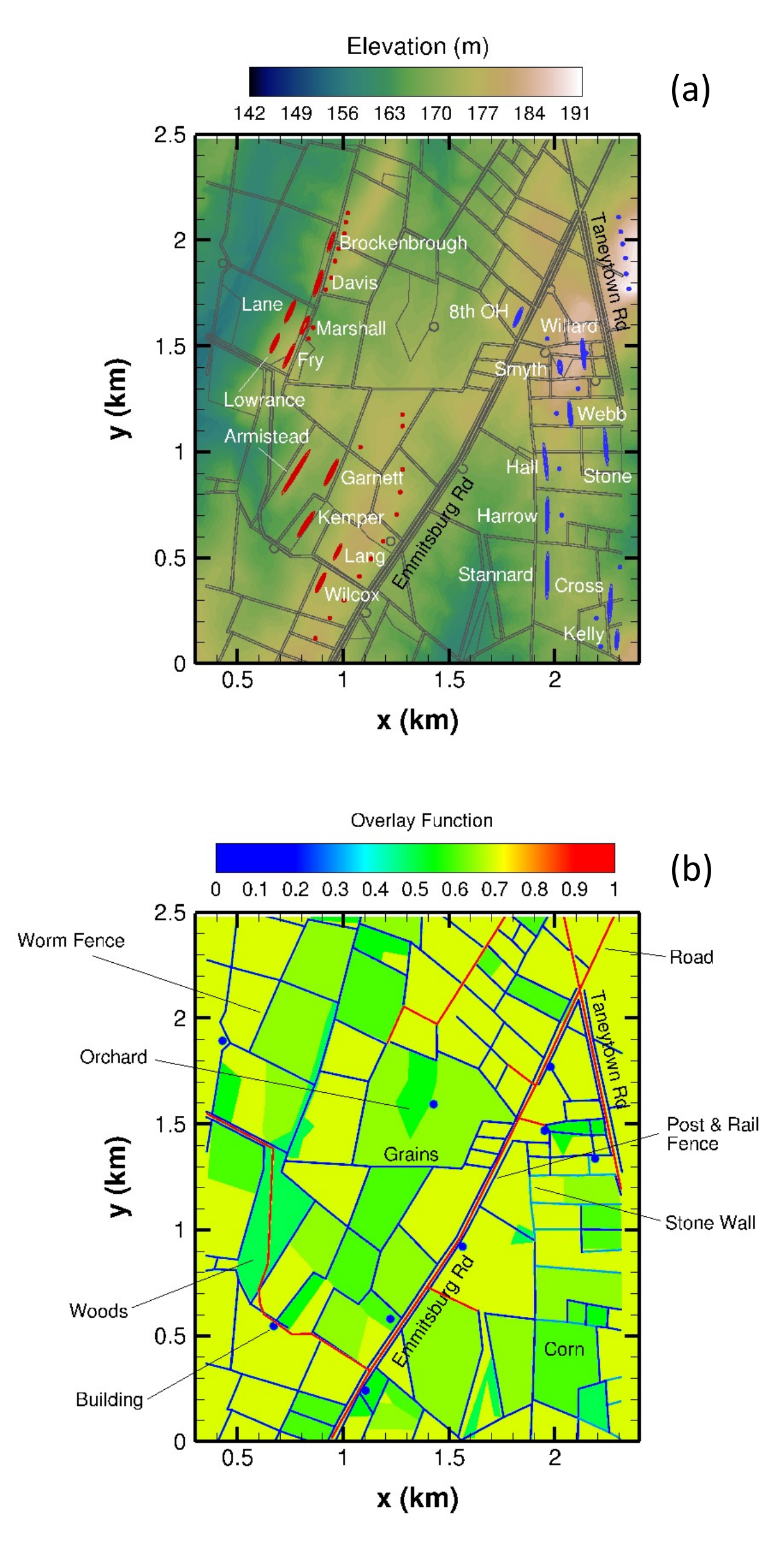}
  \caption{Battlefield details, based on the maps of
    \citet{gottfried07} and \citet{laino14}.  (a) Initial
    configuration.  Oval contours represent densities of infantry,
    dots artillery; Confederate forces in red, Union in blue.  (b)
    Overlay layer labeled with various features.}\label{fig:map}
\end{figure}

\begin{table}
  \centering
  \caption{Brigades / regiments participating in Pickett’s Charge,
    based on the data of \citet{busey05}. \label{tbl:brigades}}
  {\tiny
    \begin{tabular}{lrrrrrrr}
      \hline
      Ref.  No. & Side & Brigade Leader & Init. Morale & Est. Strength  \\
      \hline
      1 & Union & 8 OH & 0.7 & 141  \\
      1 & Union & 126 NY & 0.7 & 150 \\
      2 & Union & Willard & 0.7 & 1030  \\
      3 & Union & Smyth & 0.7 & 828  \\
      4 & Union & Webb & 0.7 & 895  \\
      5 & Union & Hall & 0.7 & 669  \\
      6 & Union & Stone & 0.8 & 745  \\
      7 & Union & Harrow & 0.7 & 831  \\
      8 & Union & Stannard & 0.7 & 1715  \\
      9 & Union & Cross & 0.7 & 632  \\
      10 & Union & Kelly & 0.7 & 399 \\
                & & & Total & 8036 \\
      \hline
      11 & Conf. & Brockenbrough & 0.8 & 829  \\
      12 & Conf. & Lane & 0.7 & 1203  \\
      13 & Conf. & Davis & 0.8 & 1484  \\
      14 & Conf. & Lowrance (Scales) & 0.7 & 879  \\
      15 & Conf. & Marshall (Pettigrew) & 0.8 & 1495  \\
      16 & Conf. & Fry (Archer) & 0.8 & 739  \\
      17 & Conf. & Armistead & 1.0 & 1223  \\
      18 & Conf. & Garnett & 1.0 & 824  \\
      19 & Conf. & Kemper & 1.0 & 1163  \\
      20 & Conf. & Lang (Perry) & 0.7 & 437  \\
      21 & Conf. & Wilcox & 0.7 & 1205 \\
                & & &  Total & 11481\\
      \hline
    \end{tabular}
  }
\end{table}

\begin{table}\centering
  \caption{Artillery units participating in Pickett’s Charge, based on
    the data of \citet{busey05}. \label{tbl:batteries}}
  {\tiny
  \begin{tabular}{lrrrr}
    \hline
    Ref. No. & Side & Battery Leader & Guns \\
    \hline
    22 & Union & Bancroft / Wilkeson & 6 \\
    23 & Union & Mason / Eakin & 6 \\
    24 & Union & Edgell & 4 \\
    25 & Union & Hill & 4 \\
    26 & Union & Norton & 6 \\
    27 & Union & McCartney & 6 \\
    28 & Union & Woodruff & 6 \\
    29 & Union & Milton / Bigelow & 6 \\
    30 & Union & Arnold & 6 \\
    31 & Union & Cushing & 6 \\
    32 & Union & Brown / Perrin & 6 \\
    33 & Union & Rorty & 4 \\
    34 & Union & Daniels & 6 \\
    34 & Union & Thomas & 6 \\
    35 & Union & Hart & 4 \\
    35 & Union & Phillips & 6 \\
    35 & Union & Thompson & 5 \\
    36 & Union & Rank & 2 \\
             & & Total & 95 \\
    \hline
    37 & Conf. & Brander & 4 \\
    38 & Conf. & McGraw & 4 \\
    39 & Conf. & Brunson / Zimmerman & 4 \\
    40 & Conf. & Johnston & 4 \\
    41 & Conf. & Marye & 4 \\
    42 & Conf. & Ross & 6 \\
    43 & Conf. & Wingfield & 5 \\
    44 & Conf. & Graham & 4 \\
    45 & Conf. & Wyatt & 4 \\
    46 & Conf. & Brooke & 4 \\
    47 & Conf. & Ward & 4 \\
    48 & Conf. & Woolfolk & 4 \\
    49 & Conf. & Blount & 4 \\
    50 & Conf. & Caskie & 4 \\
    51 & Conf. & Macon & 4 \\
    52 & Conf. & Stribling & 6 \\
    53 & Conf. & Richardson & 3 \\
    54 & Conf. & Norcom & 3 \\
    55 & Conf. & Miller & 3 \\
    56 & Conf. & Taylor & 4 \\
    57 & Conf. & Gilbert & 4 \\
             & & Total & 86 \\
    \hline
  \end{tabular}
  }
\end{table}

\begin{table}\centering
  \caption{Values selected for the overlay layer for terrain
    features. \label{tbl:terrain_features}}
  {\tiny
    \begin{tabular}{lrrr}
      \hline
      Feature & Maximum $T(\boldsymbol{x})$ & Characteristic Scale (m) \\
      \hline
      Woods & 0.50 & --- \\
      Orchard & 0.55 & --- \\
      Corn & 0.60 & --- \\
      Grain & 0.65 & --- \\
      Building & 0.05 & 20.0 \\
      Post and rail fence & 0.05 & 10.0 \\
      Worm fence & 0.10 & 10.0 \\
      Stone wall & 0.30 & 10.0 \\
      Road & 1.00 & 10.0 \\
      \hline
    \end{tabular}
  }
\end{table}

The present calculations are intended to represent the events of the
afternoon of July~3, 1863. The situation is illustrated by the
historical Gettysburg maps of Figure~\ref{fig:map}, and the units
involved are summarized in Tables~\ref{tbl:brigades}--\ref{tbl:batteries}.

The initial configuration of units is presented with labels in
Figure~\ref{fig:map}a.  The color contours show the local elevation
and the dark contour lines represent the overlay layer.  Confederate
forces are shown in red, Union in blue.  The oval contours are
brigades or regiments; the red dots are artillery units.

On the Confederate lines, the Heth / Pettigrew division
(Brockenbrough, Davis, Marshall, and Fry brigades) is initially
stationed approximately along the crest of Seminary Ridge. Pender’s
division (Lane and Lowrance) lies behind them. Pickett’s division
(Armistead, Garnett, and Kemper) stands to the south, and Anderson’s
division (Lang and Wilcox) stands close by. In front of the infantry,
several artillery positions form two lines.

The Union forces occupy a defensive position along a north-south line
along Cemetery Ridge. The combined 8~Ohio / 126~New York unit lies to
the west of Emmitsburg Road. Hays’ division (Willard and Smyth) hold
the northern end of the line. In the middle is Gibbon’s division
(Webb, Hall, and Harrow) and the Doubleday / Rowley division (Stone
and Stannard). To the south lies Caldwell’s division (Cross and
Kelly).

Figure~\ref{fig:map}b shows the elevation and the terrain overlay
layer used to represent the effects of fences, roads, buildings, and
local vegetation.  Conditions for the map overlay layer were based on
the maps provided in \citet{gottfried07} and \citet{laino14}. These
historical conditions are relatively well-known because of the
availability of the \citet{warren1869} Map, created through a survey
conducted a few years after the end of the US Civil War. In processing
the map data, regions of vegetation (corn, grain, orchard, woods) were
defined by polygons, and the overlay function value was taken to be
constant inside each polygon.

Buildings were characterized by a point; the overlay function was
defined by a central peak with a decaying distribution (error
function) away from the center with a characteristic scale.  These
appear as blue dots in Figure~\ref{fig:map}b. An example is the dot
located approximately at the coordinates (1.5~km, 1.6~km) on the map;
this represents the house and barn of the Bliss Farm.

Similarly, roads and various kinds of fences (post and rail, stone,
worm) were defined by line segments with decay in $T(\boldsymbol{x} )$
with distance from the line.  The values selected for the overlay
function for various terrain features are listed in
Table~\ref{tbl:terrain_features}. The strong fences on either side of
Emmitsburg Road were significant obstacles during the historical
battle.

The overlay layer was smoothed to obtain numerical stability; abrupt
jumps in $T(\boldsymbol{x} )$ led to numerical oscillations. For a
$384 \times 309$ numerical grid (8~m resolution), three passes of a
nine-point Laplacian averaging stencil provided sufficient smoothness
without excessive distortion of the map.

A fundamental input to the model is the number of soldiers and
artillery pieces participating in the fighting related to Pickett's
Charge.  \citet{busey05} have published a comprehensive tabulation of
the forces present at the Battle of Gettysburg.  Very accurate numbers
are available for June~30, when both sides carried out a routine roll
call.  Good figures of the forces engaged and lost (killed, wounded,
captured, or missing) have been assembled for the overall three-day
battle from accounts made in the aftermath. Nonetheless, the forces
available on the afternoon of July~3, after two days of fighting, are
unknown, and the numbers must be estimated.

The numbers quoted in the literature for the forces participating in
the fighting related to Pickett's Charge vary.  In a review of the
literature, \citet{armstrong15} found estimates (in round numbers) of
5800--8000 Union and 10500--13000 Confederate infantry engaged, with,
respectively, casualties of 1300--2300 and 5900--6500 for each side.

For the present calculations, we used figures from \citet{busey05},
with additional input from \citet{gottfried07,gottfried12},
\citet{hessler15}, and \citet{laino14} to judge which units
participated.  A force estimate was made for each unit at the start of
the Pickett’s Charge event, based on known forces and losses for the
overall battle, and history of combat on the previous two days of the
battle.

Table~\ref{tbl:brigades} provides details of the brigades and certain
regiments participating in Pickett’s Charge.  The first column
indicates the reference number used in our modeling, roughly
corresponding to position north to south on the map in
Figure~\ref{fig:map}. The next column provides the brigade leader, to
identify the unit. An exception is the first unit, which represents a
combination of parts of the 8~Ohio and 126~New York regiments that
were positioned to the west of Emmitsburg Road, apart from the main
Union lines. The table also provides an associated initial morale
score $M_i^0$, based on historical information for the date of last
combat for that unit. We assume that a unit that had engaged in recent
combat would have lower morale.

Again, \citet{busey05} provide accurate numbers for forces present at
a roll call on June~30, 1863, and good approximations of engaged
forces on the first day of the Gettysburg battle and of losses
sustained in the whole conflict.  The forces that took part in the
Pickett’s Charge scenario must be estimated.  The final column in
Table~\ref{tbl:brigades} is our estimate of those forces.

The corresponding data for artillery batteries are presented in
Table~\ref{tbl:batteries}. The artillery units are identified by
battery leader. Some of the Union batteries (Units~34 and 35) have
been aggregated for simplicity. Again, the numbers were taken from
\citet{busey05}. The selection of batteries that participated in the
Pickett’s Charge scenario is based on \citet{hessler15} and on our own
judgement.

The present approach does not require exact figures for the forces
engaged in the scenario. It is sufficient to have a good
approximation.  In our methodology, that approximation provides the
baseline for a study of the sensitivity of the outcome on the
parameters used as input in the model. (See the discussion associated
with Figure~\ref{fig:statistical}.)

According to the maps presented in \citet{gottfried12} and
\citet{laino14} all units were in formation by 13:00 on July~3. By
14:30, the brigades commanded by Brockenbrough, Davis, Marshall, and
Fry were marching toward the northern end of the Union lines, and the
unit composed of the 8~Ohio and~126 New York regiments was swinging
out to flank the Confederates. Meanwhile, Kemper and Garnett were
taking their brigades toward the southern part of the Union lines. By
14:50, Brockenbrough was withdrawing, but Lane, Lowrance, and
Armistead had begun to advance. Kemper and Garnett were shifting
northward. On the Union side, Stannard was beginning a maneuver to
flank the Confederate attack. At 15:15, most of the Confederate units
had crossed Emmitsburg Road; only Lane and Lowrance remained to the
west. Later, at 15:30, the Confederate charge was losing its momentum,
and many units were retreating. By 15:50 Lang and Wilcox had made a
late crossing of Emmitsburg Road for an attack; the rest of the
Confederates were in full retreat.

Some of the motion of the Confederate attackers may be hard to
reconstruct. \citet{reardon97} notes (p.~21) historical accounts of
“the unusual indirect approach march that the Virginians [Pickett’s
division] employed,” and quotes a survivor as stating that they “moved
alternately by the front \& by the left flank under a most deadly fire
of infantry and artillery.” The present model does not attempt to
match this apparent staircase path; diagonal straight lines with
appropriate waypoints are employed.

\subsection{Army Level of Aggregation}
\label{subsec:armies}

\begin{figure*}
  \centering
  \includegraphics[width=0.95\linewidth]{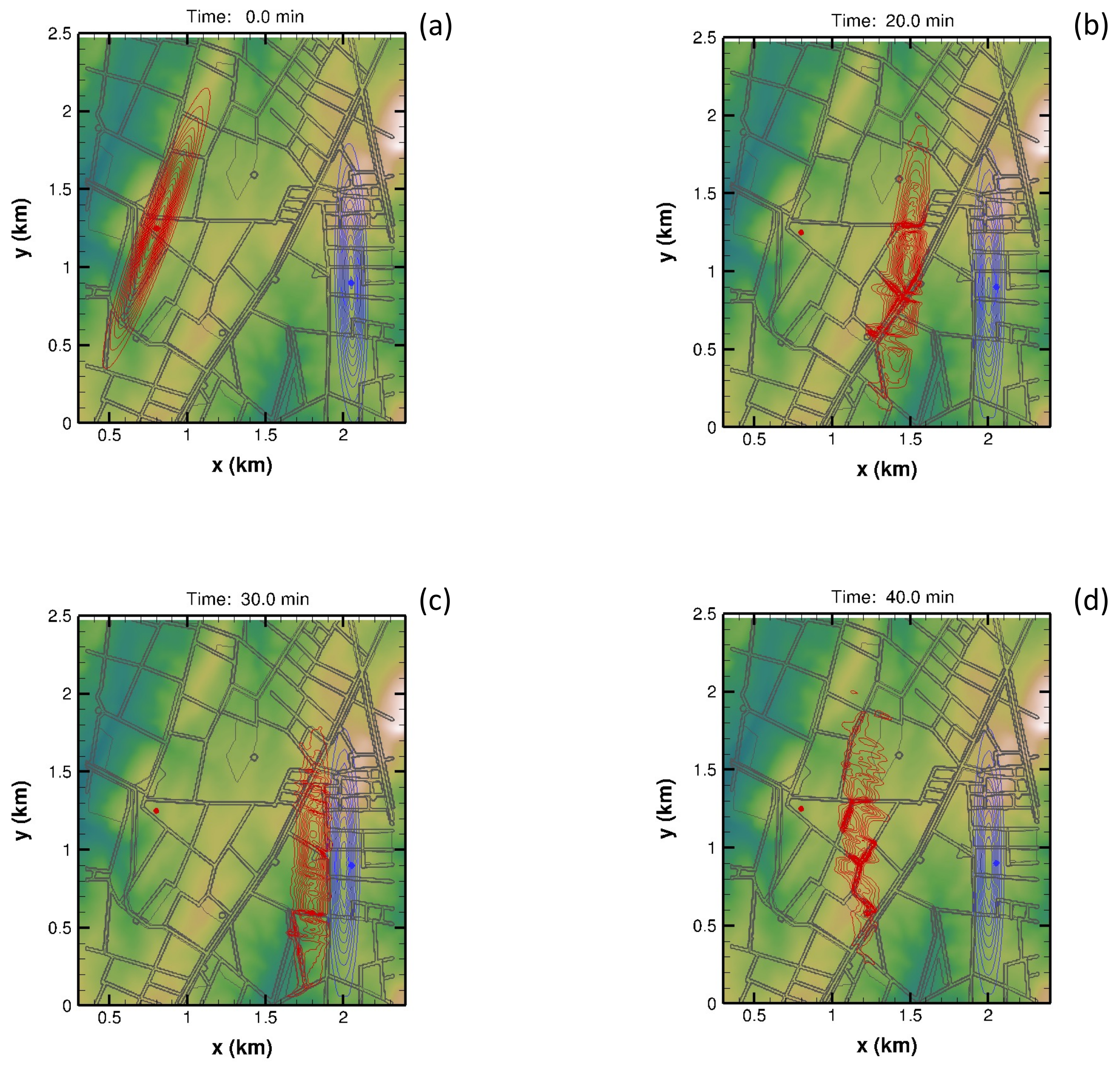}
  \caption{Results at the army level of aggregation.
    Oval contours represent densities of infantry,
    dots aggregated artillery; Confederate forces in red, Union in blue.
    (a) Initial state, 0.0~min.
    (b) Red crossing Emmitsburg Road, 20.0~min.
    (c) Initial contact, 30.0~min.
    (d) Red retreats on suffering heavy casualties, 40.0~min.
    Video available at \protect \url{https://engineering.purdue.edu/~jpoggie/battle_flow_model/armies.mp4}
    \label{fig:army_aggregation}}
\end{figure*}

\begin{figure}
  \centering
  \includegraphics[width=0.5\linewidth]{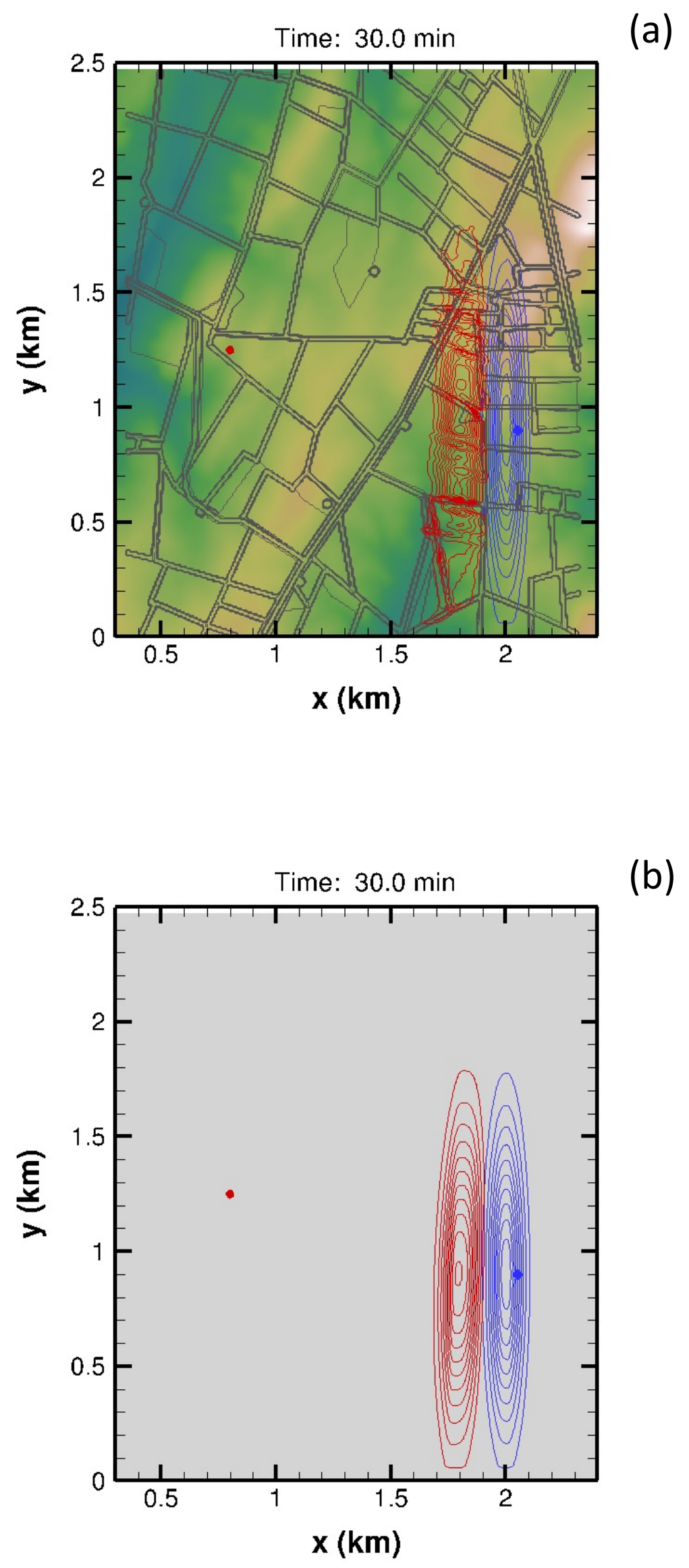}
  \caption{Effect of terrain at the army level of aggregation.
    Oval contours represent densities of infantry,
    dots aggregated artillery; Confederate forces in red, Union in blue.
    (a) Elevation and terrain effects included.
    (b) Perfectly flat landscape.
    Video available at \protect \url{https://engineering.purdue.edu/~jpoggie/battle_flow_model/armies_no_terrain.mp4}
    \label{fig:flat_landscape}}
\end{figure}

As an initial demonstration of the mathematical model, we consider the
Pickett’s Charge scenario at an army level of aggregation. There are
four entities in this version of the model: Union infantry, Union
artillery, Confederate infantry, and Confederate artillery.

The initial force strength is based on aggregation of the
brigade-level data
(Tables~\ref{tbl:brigades}--\ref{tbl:batteries}). At the start of the
battle simulation, the Union army had 8036 infantrymen and 95
artillery pieces, whereas the Confederate army had 11481 infantrymen
and 86 artillery pieces.  The computational grid consisted of
$384 \times 309$ points, or 8~m map resolution. The time step was
1.0~s, with a total run of 3600 steps, or 1.0~h of simulated time.

The model parameters used for this case were slightly different than
those employed for the brigade-level simulation described in the next
section. The initial morale was taken as $M_i^0=1.0$ for all units.
The ranged fire efficiency for infantry and artillery were,
respectively, reduced to $k_{ij}^\prime=2.0$~s$^{-1}$ and
$k_{ij}^\prime = 4.0$~s$^{-1}$.

Figure~\ref{fig:army_aggregation} shows results at selected times in
the simulation. Again, the color contours are the local elevation, the
dark lines are contours of the overlay function, red contours give the
density of Confederate forces, and blue contours the density of Union
forces. The two solid dots, initially at the army centers, represent
the aggregated artillery of each side.

Figure~\ref{fig:army_aggregation}(a) shows the initial conditions
(0.0~min); the armies are seen to be spread across the two
ridges. With the start of the simulation, the Confederate army begins
marching toward the Union position. The Union forces hold their
positions until the Confederates are close, at which time they move forward
to a defensive position at the fences and stone walls that mark the
edge of the field. (This motion is used as an indicator in the model;
it is not intended to be historically accurate.)

Initially exchanging artillery and ranged musket fire, the armies come
into close contact at 30.0~min,
Figure~\ref{fig:army_aggregation}(c). At this stage, the close-in
combat model dominates the casualty rate. After taking heavy
casualties, the Confederates are in full retreat by 40.0~min,
Figure~\ref{fig:army_aggregation}(d).

The model predicts 1736 Union and 6115 Confederate casualties by the
end of the battle. These figures are well within the range of the
historical numbers of 1300--2300 Union and 5900--6500 Confederate
casualties quoted by \citet{armstrong15}.

The case was re-run with the elevation and overlay layer effects
turned off, to test the effects of these features on the outcome of
the scenario. Figure~\ref{fig:flat_landscape} compares results for the
two cases near the time of army contact in the simulation. As might be
expected, the terrain effects are seen to produce the wrinkling of the
unit density contours that is not present for the perfectly flat
terrain case. This prediction is an important feature of the
continuous flow model; to capture the same effect in a discrete model
would require a very large number of agents.

A more striking result was a change in the total number of casualties
in the absence of terrain, a total of about 1945 Union and 5819
Confederate.  The Union forces take about 200 more casualties, and the
Confederates 300 fewer, under the conditions of perfectly flat
terrain.  The terrain effects are seen to slow the pace of battle, and
they favor the defenders. With terrain effects included, the slow
progress of the Confederate troops toward the Union lines exposes the
attacking Confederates to heavy ranged fire for an extended period.

\subsection{Brigade Level of Aggregation}
\label{subsec:brigades}

\begin{figure*}
  \centering
  \includegraphics[width=0.95\linewidth]{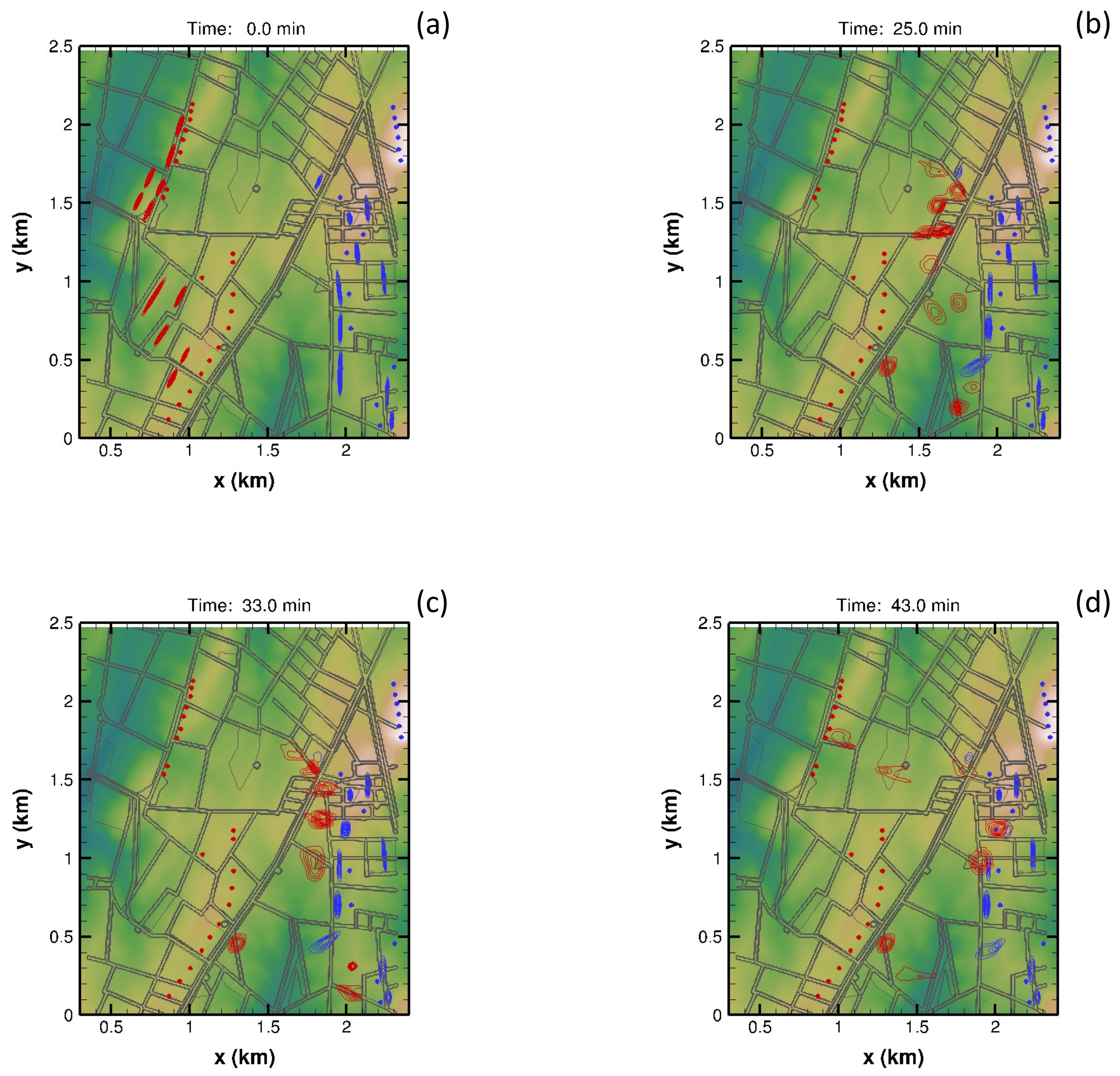}
  \caption{Results at the brigade level of aggregation (Case~0).
    Oval contours represent densities of infantry,
    dots artillery; Confederate forces in red, Union in blue.
    (a) Initial state, 0.0~min.
    (b) Initial contact, 25.0~min.
    (c) Heavy combat near The Angle, 33.0~min.
    (d) Red retreats after suffering heavy casualties, 43.0~min.
    Video available at \protect \url{https://engineering.purdue.edu/~jpoggie/battle_flow_model/brigades_case0.mp4}
    \label{fig:case0}}
\end{figure*}

The scenario for the main set of calculations was carried out at the
brigade level of aggregation.  The computational grid, time step, and
simulation duration were the same as those used for the army-level
simulation.  Selected results are shown in Figure~\ref{fig:case0}.
The initial conditions are shown in the top left subfigure at time
0.0~min.  The Confederates begin their march eastward at the start of
the simulation. At about the 18~min mark, the 8~Ohio / 126~New York
unit and Stannard's brigade swing around to flank the charging
Confederates.  At 25.0~min (top right in Figure~\ref{fig:case0}), a
group of Confederates has developed a force concentration on the
western side of Emmitsburg Road. Pickett's division, having been
delayed crossing the road, heads northeast toward The Angle.  At
33.0~min (bottom left in Figure~\ref{fig:case0}), some of the
Confederates press their attack at The Angle, but other units have
taken such casualties that they begin to withdraw. By~43.0 min, many
of the Confederate brigades are in full retreat (bottom right in
Figure~\ref{fig:case0}).

For this baseline scenario, the model predicts 2534 Union and 3620
Confederate casualties by the end of the battle, slightly out of the
range of the historical figures collected by \citet{armstrong15}. The
model parameters could be adjusted to make the predictions fall in the
middle of the estimates of the historical figures, but our approach is
to instead address variation of the model parameters in a systematic
manner.

\begin{figure*}
  \centering
  \includegraphics[width=0.95\linewidth]{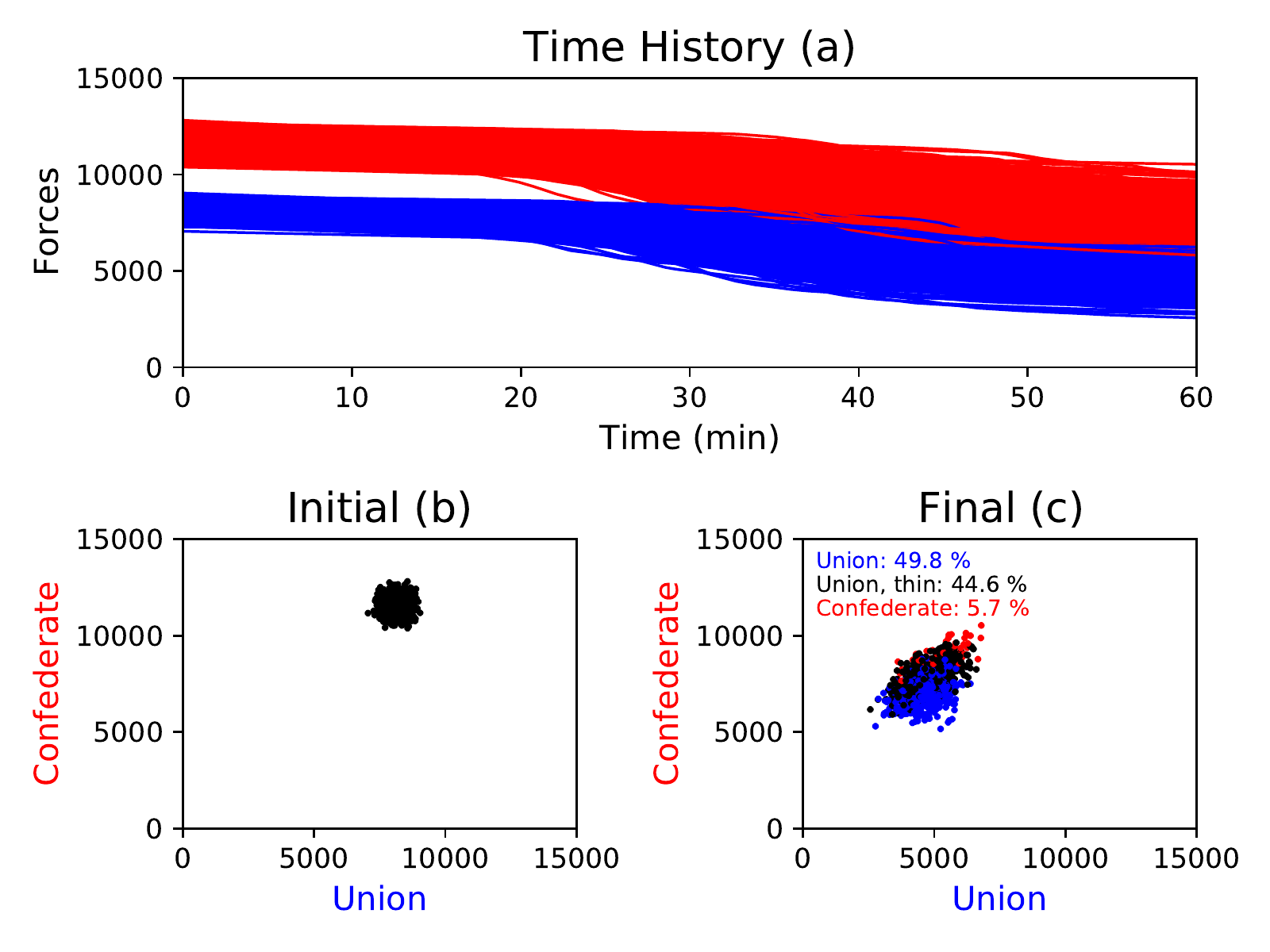}
  \caption{Statistical analysis of 1000 cases with randomized initial
    conditions and model parameters. (a) Time histories of total
    forces for each case. (b) Initial forces for each case. (c) Final
    forces for each case.}\label{fig:statistical}
\end{figure*}

To assess the sensitivity of the model to changes in the initial
conditions and model parameters, a statistical analysis of the outcome
for randomized inputs was carried out.  As with any historical event,
there is a level of uncertainty in the historical details of the
Pickett’s Charge scenario. Furthermore, as noted above, even
eyewitness accounts of the battle have been colored by the personal
and political interests of the survivors \citep{reardon97}. All of the
inputs to a model of the battle must thus be considered to lie within
an uncertainty band that may be larger than one would hope.  Both the
value and validity of the results are measured against the
uncertainties in the literature.

To generate variations on the baseline course of events (Case~0), a
program was written in the Python language to generate randomized
inputs. The algorithm employed the PCG-64 pseudo-random number
generator. For repeatability, a set of 1000 cases was generated using
a seed based on the case number (Cases~1--1000). All significant
initial conditions and model parameters were randomized, as follows.

Initial positions and goal positions were randomized using a normal
distribution with a standard deviation of 100~m. (Thus, about 95\% of
the randomly generated values lie within $\pm 200$~m of the baseline
positions.) Similarly, the initial and final bearings were randomized
with a standard deviation of $10^\circ$. The parameters characterizing
the initial and final shapes of the formation were randomized using a
uniform distribution ranging between 80\% and 120\% of the baseline
figure. A similar approach was applied to the morale parameters,
initial forces, and ordered marching speed.

To test the effect of strong variations in the effect of the
parameters characterizing combat (close-in combat factor, ranged-fire
combat factor, and weapon range), these parameters were perturbed by
factors of two, that is, with a uniform distribution between 50\% and
200\% of the baseline value. The changes in combat effectiveness were
applied evenly across all units; under this perturbation, no unit
gained an advantage or disadvantage relative to the others.

The results are shown in Figure~\ref{fig:statistical}. Time histories
of the total Union (blue lines) and Confederate (red lines) forces (an
ensemble forecast) are shown in Figure~\ref{fig:statistical}(a) for
Cases 0--1000. The results reflect slow losses from artillery fire
during the initial charge, followed by more rapid losses beginning at
about the 25~min mark when the units close to rifle range.

Figure~\ref{fig:statistical}(b) presents a plot of initial Union
versus Confederate forces; this is a kind of slice at time zero of the
time-history plot. Each dot represents the initial condition of a
particular case. The perturbation of $\pm 20$\% in the initial forces
is apparent here.

The key results are presented in Figure~\ref{fig:statistical}(c),
which shows Union versus Confederate forces at the end of the 60~min
of simulated time. Again, this is a kind of slice of the time-history
plot at the end of the battle, and each dot represents the end state
of one of the cases. The baseline Case~0 results in about 5600 Union
and 7900 Confederate survivors (about 2500 and 3600 casualties,
respectively). With the variation in the results under the strong
perturbation of the model inputs, the number of survivors lies in a
range of about $\pm 2500$ for each side, the scatter in
Figure~\ref{fig:statistical}(c). Given these high levels of
casualties, these results are hard to interpret by themselves.

To clarify the outcome and facilitate interpretation, a set of victory
criteria was established based on the number of Confederate soldiers
in retreat at the end of the 60 min of simulated time. If 50\% or more
of the surviving Confederate forces were in retreat, the result was
considered a conclusive Union victory. If 10\% or less of the total
Confederate forces were in retreat, the result was considered a
Confederate victory. Results in between these thresholds were
considered a thin Union victory.

In Figure~\ref{fig:statistical}(c), the outcomes are color coded by
these victory criteria. The lack of possible outcomes in favor of the
Confederates is immediately apparent. Under these criteria, about half
(49.8\%) of the scenarios resulted in a conclusive Union
victory. Another, almost equal, percentage (44.6\%) of outcomes
resulted in a thin Union victory. Only 5.7\% of the cases could be
considered Confederate victories. In the context of our model, then, the
Confederate forces had little chance of victory in the Pickett’s
Charge battle.

\section{Discussion}

A continuous flow model of infantry behavior \citep{poggie20} has been
developed in sufficient detail that it can now be applied to a
realistic simulation of a historical battle.  The model is based on
conservation of individuals and tracking of subunit identity.
Extensions to the model include artillery units, an overlay layer to
represent different kinds of terrain on the map, improvements to
movement and ranged fire, a list of orders for each unit, unit morale,
and checks for retreat and pressing an attack.

Pickett’s charge during the 1863 Battle of Gettysburg, Pennsylvania in
the U.S. Civil War was chosen as an initial application of the model.
This scenario is a good test of the current mathematical model because
many modern military tactics were employed, in a context where the
action took place on foot or horseback. Furthermore, detailed maps and
figures for troop numbers are available for the historical battle.

An initial simulation at the army level of aggregation was carried out
with and without terrain effects.  Elevation, buildings, vegetation,
and particularly fences slowed the advance of the Confederate attack
when terrain was included.  The resulting differences in troop density
and position had consequences for both close-in and ranged combat.
With terrain effects included, the slow progress of the Confederate
troops toward the Union lines exposed the attacking Confederates to
heavy ranged fire for an extended period, and resulted in higher
Confederate casualties.

The results illustrate the strength of the continuous flow model in
representing the effect of terrain on infantry movement.  To capture
the same effect in a discrete model would require an unworkably large
number of agents.  Compared to a discrete agent model with the same
number of units, we judge that the continuous flow model better
captures the interaction of the forces with the terrain and each
other.

The main case study was a brigade-level simulation of the
conflict. This was intended to be as faithful as possible to the
historical events, and that aim appears to be achieved. The main
source of asymmetry in the numbers of casualties was the inability of
the Confederate forces to use effective ranged fire while they were
moving. Again, the terrain slowed the pace of battle and favored the
defenders, exposing the attackers to heavy ranged fire for an extended
period.

A statistical analysis of possible outcomes for randomized
perturbations of the baseline brigade-level scenario was carried
out. These perturbations encompassed all the initial conditions and
all the significant model parameters characterizing combat. Using the
number of Confederate units in retreat at the end of the 60 min
simulation as a threshold, it was found that only 6\% of the scenarios
resulted in an outcome that could be considered a Confederate victory.

Our results differ somewhat from those \citet{armstrong15} obtained
using a Lanchester model.  Those authors concluded that the
Confederates could plausibly have taken the Union position under a
different tactical approach.

In the context of the present mathematical model, which better
represents the spatial evolution of the conflict, no reasonable
concentration of Confederate forces had sufficient numerical advantage
to break through and hold a section of the Union lines in a militarily
significant manner.  Given the limited advantage in numbers held by
the Confederates, over-concentration of their forces (say on The
Angle) would leave their flanks open to Union counterattack. They must
assign at least some forces to each part of the Union lines. Given
that situation, it is almost impossible for them to achieve the 3:1
local numerical advantage that is held to be required for
victory \cite{davis95,kress99}.

\section*{Acknowledgements}

The authors are indebted to Matthew Ellis for help with the
topographical maps. Collin Tofts converted map data to the overlay
layer. Matthew Konkoly provided many valuable suggestions for
improving the combat model. Carol Reardon provided us with an
invaluable overview of the historical battle on site, and an
appreciation for the high uncertainty about many of the historical
details. Anna Creese researched details and provided helpful
discussion. Computational resources were provided by Purdue
University’s Rosen Center for Advanced Computing \citep{hacker14}.


\begin{thebibliography}{29}
\providecommand{\natexlab}[1]{#1}
\providecommand{\url}[1]{\texttt{#1}}
\expandafter\ifx\csname urlstyle\endcsname\relax
  \providecommand{\doi}[1]{doi: #1}\else
  \providecommand{\doi}{doi: \begingroup \urlstyle{rm}\Url}\fi

\bibitem[Lighthill and Whitham(1955)]{lighthill55}
M.~J. Lighthill and G.~B. Whitham.
\newblock On kinematic waves. {II}. {A} theory of traffic flow on long crowded
  roads.
\newblock \emph{Proceedings of the Royal Society of London. Series A,
  Mathematical and Physical Sciences}, 229\penalty0 (1178):\penalty0 317--345,
  1955.
\newblock URL \url{https://www.jstor.org/stable/99769}.

\bibitem[Protopopescu et~al.(1989)Protopopescu, Santoro, and
  Dockery]{protopopescu89}
V.~Protopopescu, R.~T. Santoro, and J.~Dockery.
\newblock Combat modeling with partial differential equations.
\newblock \emph{European Journal of Operational Research}, 38\penalty0
  (2):\penalty0 178--183, 1989.
\newblock URL \url{https://doi.org/10.1016/0377-2217(89)90102-1}.

\bibitem[Fields(1993)]{fields93}
M.~A. Fields.
\newblock Modeling large scale troop movement using reaction diffusion
  equations.
\newblock Technical Report ARL-TR-200, Army Research Laboratory, September
  1993.

\bibitem[Hughes(2002)]{hughes02}
R.~L. Hughes.
\newblock A continuum theory for the flow of pedestrians.
\newblock \emph{Transportation Research Part B: Methodological}, 36\penalty0
  (6):\penalty0 507--535, 2002.
\newblock URL \url{https://doi.org/10.1016/S0191-2615(01)00015-7}.

\bibitem[Clements and Hughes(2004)]{clements04}
R.~R. Clements and R.~L. Hughes.
\newblock Mathematical modelling of a mediaeval battle: The {B}attle of
  {A}gincourt, 1415.
\newblock \emph{Mathematics and Computers in Simulation}, 64\penalty0
  (2):\penalty0 259--269, 2004.
\newblock URL \url{https://doi.org/10.1016/j.matcom.2003.09.019}.

\bibitem[Spradlin and Spradlin(2007)]{spradlin07}
C.~Spradlin and G.~Spradlin.
\newblock Lanchester’s equations in three dimensions.
\newblock \emph{Computers and Mathematics with Applications}, 53\penalty0
  (7):\penalty0 999--1011, 2007.
\newblock URL \url{https://doi.org/10.1016/j.camwa.2007.01.013}.

\bibitem[Lanchester(1914)]{lanchester14}
F.~W. Lanchester.
\newblock Aircraft in warfare: The dawn of the {F}ourth {A}rm --- {N}o. {V}:
  The principle of concentration.
\newblock \emph{Engineering}, 98:\penalty0 422--423, 1914.

\bibitem[Keane(2011)]{keane11}
T.~Keane.
\newblock Combat modeling with partial differential equations.
\newblock \emph{Applied Mathematical Modelling}, 35\penalty0 (6):\penalty0
  2723--2735, 2011.
\newblock URL \url{https://doi.org/10.1016/j.apm.2010.11.057}.

\bibitem[Gonz{\'a}lez and Villena(2011)]{gonzalez11}
E.~Gonz{\'a}lez and M.~Villena.
\newblock Spatial {L}anchester models.
\newblock \emph{European Journal of Operational Research}, 210:\penalty0
  706--715, 2011.
\newblock URL \url{https://doi.org/10.1016/j.ejor.2010.11.009}.

\bibitem[Poggie et~al.(2020)Poggie, Matei, and Kirchubel]{poggie20}
J.~Poggie, S.~A. Matei, and R.~Kirchubel.
\newblock Simulating military conflict with a continuous flow model.
\newblock \emph{Journal of the Operational Research Society}, 2020.
\newblock URL \url{https://doi.org/10.1080/01605682.2020.1825017}.

\bibitem[Gottfried(2007)]{gottfried07}
B.~M. Gottfried.
\newblock \emph{The Maps of {G}ettysburg: An Atlas of the {G}ettysburg
  Campaign, June 3 – July 13, 1863}.
\newblock Savas Beatie, New York, 2007.

\bibitem[Laino(2014)]{laino14}
P.~Laino.
\newblock \emph{{G}ettysburg Campaign Atlas}.
\newblock Gettysburg Publishing, Trumbull CT, 3rd edition, 2014.

\bibitem[Busey and Martin(2005)]{busey05}
J.~W. Busey and D.~G. Martin.
\newblock \emph{Regimental Strengths and Losses at {G}ettysburg}.
\newblock Longstreet House, Hightstown NJ, 4th edition, 2005.

\bibitem[Weigley(2000)]{weigley00}
R.~Weigley.
\newblock \emph{A Great {C}ivil {W}ar: A Military and Political History,
  1861-1865}.
\newblock Indiana University, Bloomington, 2000.

\bibitem[Gallagher(1994)]{gallagher94}
G.~W. Gallagher, editor.
\newblock \emph{The Third Day at {G}ettysburg and Beyond}.
\newblock The University of North Carolina Press, Chapel Hill NC, 1994.

\bibitem[Reardon and Vossler(2016)]{reardon16}
C.~Reardon and T.~Vossler.
\newblock \emph{The {G}ettysburg Campaign June-July 1863}.
\newblock St. John’s Press, Alexandria VA, 2016.

\bibitem[Tillman and Engle(1996)]{tillman96}
M.~E. Tillman and C.~B. Engle, III.
\newblock An historical reenactment of the {B}attle of {G}ettysburg on {J}anus
  ({A}rmy).
\newblock \emph{Mathematical and Computer Modelling}, 23\penalty0
  (1/2):\penalty0 1--8, 1996.
\newblock URL \url{https://doi.org/10.1016/0895-7177(95)00210-3}.

\bibitem[Armstrong and Sodergren(2015)]{armstrong15}
M.~J. Armstrong and S.~E. Sodergren.
\newblock Refighting {P}ickett’s charge: Mathematical modeling of the {C}ivil
  {W}ar battlefield.
\newblock \emph{Social Science Quarterly}, 96\penalty0 (4):\penalty0
  1153--1168, 2015.
\newblock URL \url{https://doi.org/10.1111/ssqu.12178}.

\bibitem[Greenshields et~al.(1935)Greenshields, Bibbins, Channing, and
  Miller]{greenshields35}
B.~D. Greenshields, J.~R. Bibbins, W.~S. Channing, and H.~H. Miller.
\newblock A study of traffic capacity.
\newblock \emph{Highway Research Board Proceedings}, 14\penalty0 (1):\penalty0
  448--477, 1935.

\bibitem[MacKay(2009)]{mackay09}
N.~J. MacKay.
\newblock Lanchester models for mixed forces with semi-dynamical target
  allocation.
\newblock \emph{Journal of the Operational Research Society}, 60:\penalty0
  1421--1427, 2009.
\newblock URL \url{https://doi.org/10.1057/jors.2008.97}.

\bibitem[Bracken(1995)]{bracken95}
J.~Bracken.
\newblock Lanchester models of the {A}rdennes campaign.
\newblock \emph{Naval Research Logistics}, 42\penalty0 (4):\penalty0 559--577,
  1995.
\newblock URL
  \url{https://doi.org/10.1002/1520-6750(199506)42:4<559::AID-NAV3220420405>3.0.CO;2-R}.

\bibitem[Gaitonde and Shang(1993)]{gaitonde93}
D.~Gaitonde and J.~S. Shang.
\newblock Accuracy of flux-split algorithms in high-speed viscous flows.
\newblock \emph{AIAA Journal}, 31\penalty0 (7):\penalty0 1215--1221, 1993.
\newblock URL \url{https://doi.org/10.2514/3.11755}.

\bibitem[Warren(1869)]{warren1869}
G.~K. Warren.
\newblock Battlefield of {G}ettysburg.
\newblock Map, Office of the Chief of Engineers, US Army, Washington DC, 1869.
\newblock URL \url{https://www.loc.gov/item/99448794/}.

\bibitem[Gottfried(2012)]{gottfried12}
B.~M. Gottfried.
\newblock \emph{Brigades of {G}ettysburg: The Union and Confederate Brigades at
  the {B}attle of {G}ettysburg}.
\newblock Skyhorse Publishing, New York, 2012.

\bibitem[Hessler et~al.(2015)Hessler, Motts, and Stanley]{hessler15}
J.~A. Hessler, W.~E. Motts, and S.~A. Stanley.
\newblock \emph{{P}ickett’s Charge at {G}ettysburg, A Guide to the Most
  Famous Attack in American History}.
\newblock Savas Beatie, El Dorado Hills CA, 2015.

\bibitem[Reardon(1997)]{reardon97}
C.~Reardon.
\newblock \emph{{P}ickett’s Charge in History and Memory}.
\newblock The University of North Carolina Press, Chapel Hill NC, 1997.

\bibitem[Davis(1995)]{davis95}
P.~K. Davis.
\newblock Aggregation, disaggregation, and the 3:1 rule in ground combat.
\newblock Report MR-638-AF/A/OSD, RAND, Santa Monica CA, 1995.

\bibitem[Kress and Talmor(1999)]{kress99}
M.~Kress and I.~Talmor.
\newblock A new look at the 3:1 rule of combat through {M}arkov stochastic
  {L}anchester models.
\newblock \emph{Journal of the Operational Research Society}, 50:\penalty0
  733--744, 1999.
\newblock URL \url{https://doi.org/10.1057/palgrave.jors.2600758}.

\bibitem[Hacker et~al.(2014)Hacker, Yang, and McCartney]{hacker14}
T.~Hacker, B.~Yang, and G.~McCartney.
\newblock Empowering faculty: A campus cyberinfrastructure strategy for
  research communities.
\newblock \emph{Educause Review}, 2014.

\end{thebibliography}

\end{document}